\documentclass[structabstract]{aa}  

\usepackage{graphicx}
\usepackage{epstopdf}
\usepackage{txfonts}
\usepackage{natbib}

\newcommand{\doce}{\mbox{$^{12}$CO}}
\newcommand{\trece}{\mbox{$^{13}$CO}}

\newcommand{\kms}{\mbox{km~s$^{-1}$}}

\def\arcdeg{\hbox{$^\circ$}}

\begin{document}
   \title{The molecular envelope of CRL\,618: \\ A new model based on Herschel/HIFI 
   \thanks{{\it Herschel} is an ESA space observatory with science instruments provided by the
   European-led Principal Investigator consortia and with important participation from NASA.  HIFI is the 
   Heterodyne Instrument for the Far Infrared on board the Herschel space observatory.
   }
   observations.}
   \subtitle{}
   \author{R. Soria-Ruiz
          \inst{1}
          \and
          V. Bujarrabal\inst{2}
          \and
          J. Alcolea\inst{1} 
          }
   \institute{Observatorio Astron\'omico Nacional, c/ Alfonso XII 3, 28014 Madrid, Spain\\
              \email{r.soria@oan.es}
         \and
             Observatorio Astron\'omico Nacional, Ap 112, 28803 Alcal\'a de Henares, Spain\\
             }

   \date{}

 
  \abstract
   {}
   {We study the physical properties and molecular excitation of the different warm gas components
found in  the protoplanetary nebula CRL\,618. The proper study of the nebular structure and its
implications on the dynamics and kinematics of the molecular gas are of particular importance for
understanding the evolution of these objects. }
   {We revise our previous Herschel/HIFI observations, which consist of several $^{12}$CO and $^{13}$CO
lines in the far-infrared/sub-mm band in the nebula CRL\,618. These data have been
re-analyzed in detail by improving calibration, the signal-to-noise-ratio, and baseline substraction.
 Due to the high performance of Herschel,  it was possible to identify the contributions of the different
nebular components to the line profiles. Previous optical imaging and
mm-wave interferometric mapping revealed that CRL\,618 shows a complex molecular structure composed of
a large and diffuse spherical halo, a compact central core, double shells, and a fast bipolar
outflow. We have used a spatio-kinematical model to better constrain the temperature, density, and kinematics
of the molecular components probed by the improved CO observations.}
   {The $^{12}$CO and $^{13}$CO $J$=16--15, $J$=10--9, and $J$=6--5 transitions are detected in this
 source. The line profiles present a composite structure showing  spectacular wings in some cases, 
 which become dominant as the energy level increases. Our analysis of the high-energy
CO emission  with the already known low-energy $J$=2--1 and $J$=1--0 lines confirms 
that the high-velocity component, or the fast bipolar outflow, is hotter than previously estimated 
with a typical temperature of $\sim$\,300 K. This very fast component may then be an example of a 
very recent acceleration of the gas by shocks that has not yet cooled down. We also find that 
the dense central core is characterized by a very low expansion velocity, $\sim$\,5 \kms, and a strong velocity 
gradient. 
We conclude that this component is very likely to be the unaltered circumstellar layers that are lost in the 
last AGB phase, where the ejection velocity is particularly low. The physical properties of the 
other two nebular components, the diffuse halo and the double empty shell, more or less agrees 
with the estimations derived in previous models.}
   {}

   
   \titlerunning{The molecular components in CRL\,618}
   
   \maketitle
%

\section{Introduction}

Asymptotic Giant Branch (AGB) stars are pulsating red giants surrounded by spherical
and slowly expanding circumstellar envelopes (CSE) formed by their copious mass loss. In the late AGB phase,
the mass loss rate can be as high as 10$^{-4}M_{\odot}$yr$^{-1}$, and most of the stellar atmosphere is removed, 
exposing the central star. Once the mass loss ceases, the CSE expands and cools, and the effective temperature of
 the central star increases to ionize the circumstellar material. The CSE with its central AGB star
 evolves to form a planetary nebula (PN) that surrounds a blue or white dwarf. The transition between these
two stages is known as protoplanetary nebula (PPN) or post-AGB phase.  
During this PPN transition, which is believed to be very short and lasting only about 1000 yr, the nebula
experiments dramatically changes in its morphology and its physical and chemical properties.

The PPNe show fast bipolar outflows and slower components that are probably the remnants of
the ejected material in the previous AGB phase. The spectacular dynamics usually observed
in these objects is thought to be the result from the interaction between the AGB and  post-AGB winds. However, the way this
process takes place is not well understood yet. The PPNe are the ideal targets for
understanding basic and important aspects in the stellar evolution, as are the mass-loss mechanisms during 
the late AGB and post-AGB phases, and the shaping of PNe. A deep study of PPNe can be approached through multiwavelength 
observations from the visible or near-infrared ranges, which tend to select the hot and diffuse regions 
with temperatures $>$ 1000 K, to the warm and dense regions, which have a temperature range of 100
K $\le$ $T_{\rm{k}}$ $\le$ 1000 K and are traced by the far-infrared and
sub-mm observations, with cooler gas probed by mm-wave molecular
lines. Because the various components in PPNe show strong emission in molecular lines, their observation is the
best tool to study these nebulae. Thanks to these observations, the dynamics, structure, and physical
conditions in several PPNe are well known \citep [e.g.][]{bujarra01,sc04}. 

One of the best sources to study the evolutionary sequence from the AGB to the PN stages is CRL\,618. 
This is a well known PPN that initiated its post-AGB phase about 100 yr ago 
\citep{kwok84} and that is rapidly evolving to form a PN. The atomic and ionized nebula is composed of a
compact \mbox{0\farcs2 $\times$ 0\farcs4} HII region that is visible at cm- and mm-radio continuum
wavelengths \citep{wynn,kwok84,sc04,nakashima}, which is expanding at  $\sim$\,20 \kms, and of 
multiple optical lobes \citep{trammell02} with shocked gas that expands with velocities up to $\sim$\,200 \kms \citep{sc04-b}. 
Most of the circumstellar material in this source is still in molecular gas. In their 
interferometric  \doce\  $J$=2--1 maps, \citet{sc04} identified several molecular components in this source: (1)
A roughly spherical and extended halo with a mass of 0.05 $M_{\odot}$ and that is $\sim$\,20\arcsec\ in size that expands
at a low velocity of $\sim$\,17 \kms, which may be the result of the spherical mass ejection of the
central star during the AGB; (2) a fast bipolar outflow with a mass of 0.09 $M_{\odot}$ and that is 
$\lesssim$\,3$\arcsec$ in size with velocities that increase with the distance to the central star from
$\sim$\,40 to $\sim$\,340 \kms, as explained by the result of the interaction of the
AGB and post-AGB winds; (3) a compact and dense central core with a mass of 0.08 $M_{\odot}$ and a low
expansion velocity of $\lesssim$\,12 \kms, which is believed to be a more recent mass loss episode;
 and (4) another slow axial component with a mass of 0.04 $M_{\odot}$ that
is radially expanding at moderate velocities, $\sim$\,22 \kms, which may represent the entrained layers
of gas between the fast-flow and the slow AGB wind.

The most common molecular lines usually observed in PPNe are the low-$J$ rotational transitions of
\doce\ and \trece\ ($J$=2--1 and $J$=1--0). Thanks to extensive studies of these lines that are performed by means
of single-dish observations and interferometric mapping \cite[see for example,][]{bujarra01,castro10},
 important properties of PPNe have been discovered. However, they are not very useful for studying the gas at
  higher temperatures since these molecular transitions require 
 temperatures of only $\sim$\,10--20 K to be excited. Because of the crucial role of shocks in the formation and 
 evolution of PPNe, it is particularly important to observe these warmer regions.

A preliminary analysis of the Herschel/HIFI observations in CRL\,618 was presented in \citet{bujarra10}. 
Several \doce\ and \trece\ lines and emission from other molecules, like CN, HCN or H$_{2}$O, were deteced.
In this study, the authors made a preliminary nebula model to fit the $J$=16-15 \doce\ transition, concluding
that a theoretical approach including all the CO detections was necessary to better derive 
the physical properties of the different molecular components found in CRL\,618. 

We present a new calibration and analysis of several rotational lines of $^{12}$CO and $^{13}$CO  in 
CRL\,618, as previously presented in \citet{bujarra10}. Our improved observational results have been fitted by a nebula model 
based on the different molecular components known to coexist in the envelope of CRL\,618. We  also have
included  the low-$J$ lines ($J$=2--1 and $J$=1--0) of \doce\ and \trece\ in this study.
All the detected CO transitions have been simultanously modeled, leading to a 
deeper understanding of the excitation and kinematics of the molecular structures of CRL\,618.

%
%
\begin{table}
\caption{Summary of the CO lines observed in CRL\,618.}
\begin{tabular}{lr@{.}lr@{.}lcccc}
\hline 
\hline\\[-8pt]
Observed & \multicolumn{2}{c}{$\nu$} & \multicolumn{2}{c}{$T_{\rm{mb}}$} &$\sigma$ & HPBW & Cal.\\[2pt]
CO line     &\multicolumn{2}{c}{(GHz)}   & \multicolumn{2}{c}{(K)} & (K) & (\arcsec) & error\\

\hline
\hline\\[-7pt]

$^{12}$CO $J$=1--0 & 115&27  & 7&5&0.01& 22 & 20\%\\
$^{12}$CO $J$=2--1 & 230&54 &13&1&0.03& 12 & 20\%\\
$^{12}$CO $J$=6--5 & 691&47&1&1&0.01& 31 & 15\%\\
$^{12}$CO $J$=10--9 & 1151&99 &2&0&0.05& 20 &20\%\\
$^{12}$CO $J$=16--15 & 1841&34 & 2&4&0.08& 12& 30\% \\
$^{13}$CO $J$=1--0 & 110&20 & 0&5&0.01& 22 & 20\%\\
$^{13}$CO $J$=2--1 & 220&40 &2&4&0.04& 12 &20\%\\
$^{13}$CO $J$=6--5 & 661&07 &0&3&0.01& 31 &15\%\\
$^{13}$CO $J$=10--9 & 1101&35 &0&4&0.02& 20 &20\%\\
$^{13}$CO $J$=16--15 & 1760&49 &0&2&0.04& 12& 30\%\\
\hline
\end{tabular}
\label{tab1}
\end{table}

\section{Observations and data calibration}

We used the Heterodyne Instrument for the Far Infrared (HIFI) on board the Herschel Space
Observatory \citep{pil} to observe the high excitation transitions ($J$=16--15, $J$=10--9, and
$J$=6--5) of $^{12}$CO and $^{13}$CO in the PPN CRL\,618. These observations are part of the 
Guaranteed Time Key Program HIFISTARS. The preliminary version of these CO data, as  
published by \citet{bujarra10,bujarra12}, have been re-analyzed in detail by improving the
calibration and the signal-to-noise-ratio. Additional molecular lines from other species like
H$_{2}$O or HCN were also detected within the observed frequency ranges \citep [for details
see][]{bujarra10,bujarra12}, but are not discussed here.  These HIFI data are combined 
with the $J$=2--1 and $J$=1--0 results of $^{12}$CO and $^{13}$CO, as observed in this 
source by \citet{bujarra01} with the IRAM 30m telescope. Rest frequencies, main-beam  
temperatures, and calibration uncertainties of the observations are given in Table\ref{tab1}.

The HIFI operates as a DSB (double sideband) receiver. In this mode, the sky frequencies above
and below the local oscillator frequency are simultaneously observed. Then, the intermediate frequency
(IF) signal is fed into the spectrometer, which was  the HIFI wide-band spectrometer 
(WBS) in our case, an acousto-optical spectrometer that provides simultaneous coverage of the full IF band 
in the two available orthogonal receivers (H and V). 

The data were observed using the dual beam switching (DBS) mode. The HIFI internal mirror chops
between the source position and an emission-free position 3\arcmin\ away. Using
this method, the residual standing waves are expected to be minimized. This DBS
procedure worked well except for the high frequency bands 7A and 7B and especially in the V receiver,
where strong ripples were found. 

We processed the data using the standard HIFI pipeline within the HIPE software. A modified version
of the level 2 algorithm was used, yielding unaveraged time spectra but with the sub-bands stitched together.
All the observed spectra for each sub-band and receiver were checked, and those that showed 
considerable ripples were removed. Later, the data were exported using the HIPE hiClass tool to 
CLASS for further inspection, calibration, baseline removal, and final time averaging.   
When there were no significant differences between the lines detected in the H and V receivers, 
the data were systematically averaged.
Finally, the data originally calibrated in antenna temperature units were converted into 
main-beam temperatures (T$_{mb}$), using the latest available values for
the telescope and beam efficiencies \citep{roelfsema}.

\section{Line profile results}

The presence or absence of different spectral features in the profiles of the observed molecular
lines can be used to diagnose the physical properties of the diverse nebular regions found in PPNe.
If, in addition, several transitions of the same molecule are simultaneously observed, 
the overall structure of the PPNe can be studied from the regions where the CO excitation
requires temperatures of 10--20 K, which are probed by $J$=2--1 and $J$=1--0 lines,  up to the warm
regions with typical excitation temperatures of $\sim$\,800 K, as in the $J$=16--15 transitions.

\begin{figure*}[]
\centering
\includegraphics[width=0.43\textwidth]{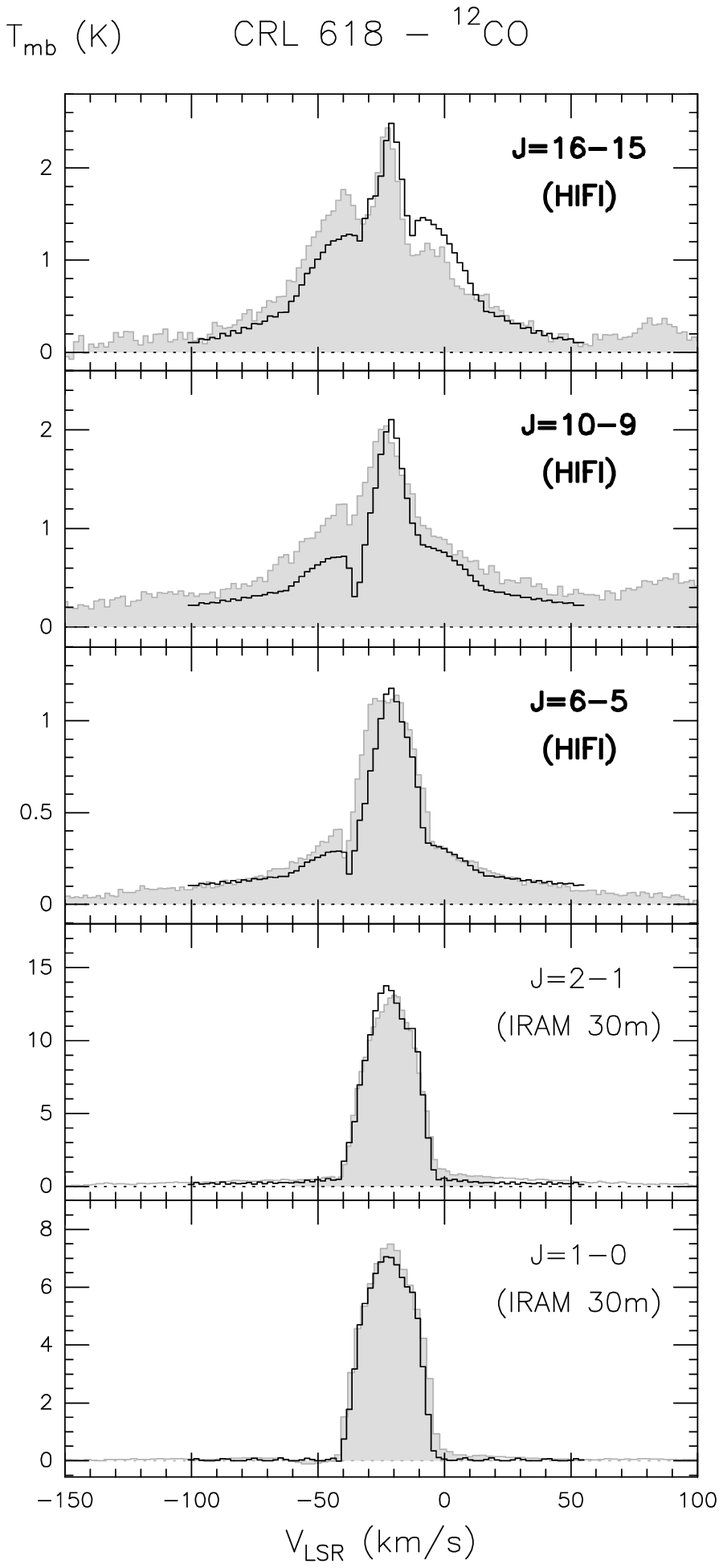}
\hspace{1cm}\includegraphics[width=0.43\textwidth]{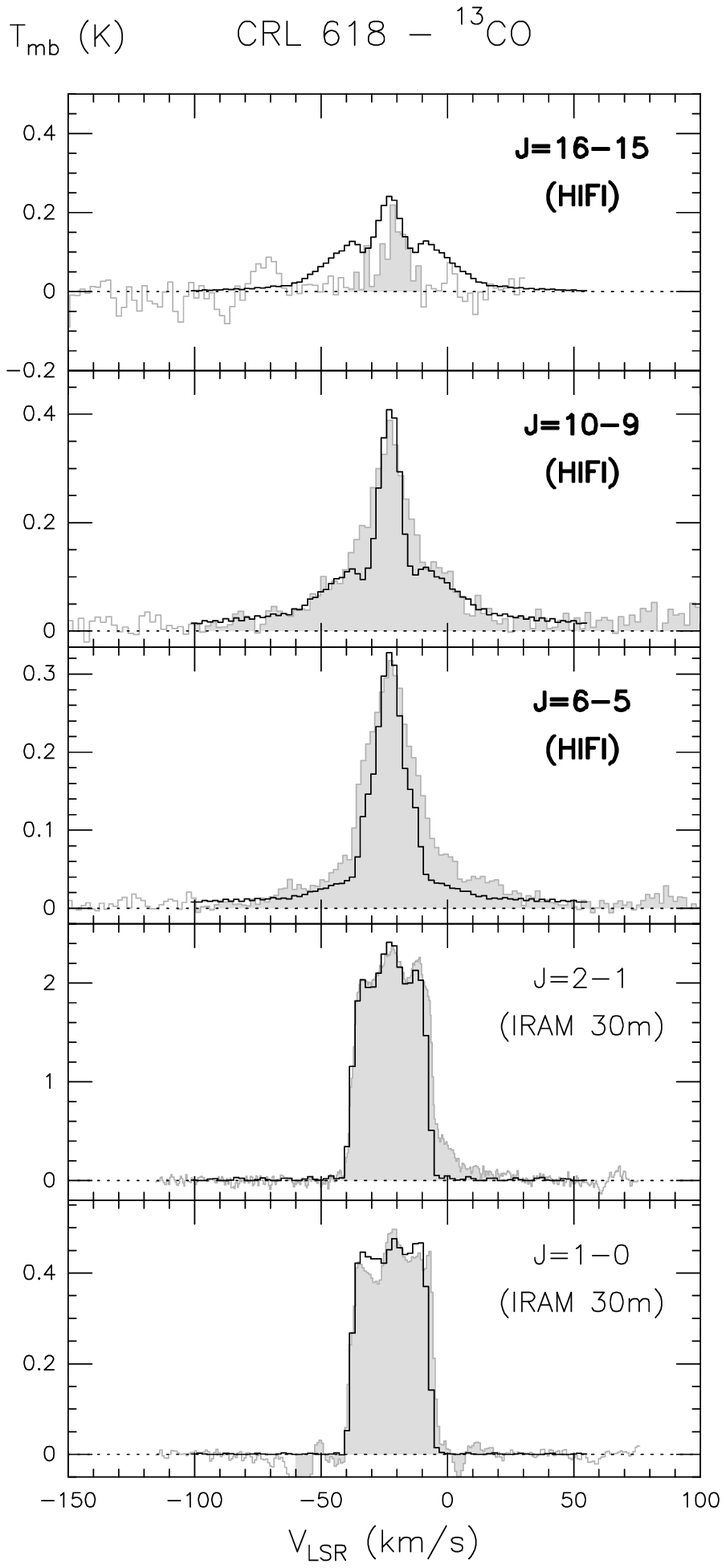}

\caption{$^{12}$CO (left) and $^{13}$CO (right) molecular lines observed in CRL\,618 (filled histogram) and model-fitting
results (solid lines). Systemic velocity of the source of $V_{sys}$=--21 \kms. We note that other
molecular lines appear within the shown spectral range, especially in the \doce\  high-$J$ profiles.}

\label{fig1}
\end{figure*}

In Fig. 1, we show the single-dish spectra of the observed $^{12}$CO transitions in CRL\,618:
$J$=16--15, 10--9, 6--5, 2--1, and 1--0. The data have been smoothed to obtain a velocity
resolution of $\sim$\,2 \kms. In general, the profiles present a composite structure with
a dominant central component (being the peak in all lines centered at the source systemic velocity,
$V_{sys}$\,=\,--21 \kms) and wide wings. From interferometric mapping, we know that the emission from
the central component corresponds to an extended ($\sim$\,20\arcsec) halo expanding at 17 \kms\ and that
the wings trace a high-velocity ($\sim$\,250 \kms) compact bipolar outflow  of about
3--4\arcsec \citep{bujarra88,neri92, bujarra01,sc04,nakashima}. We also know that additional
nebular components have been found: a central compact core surrounding the HII region, which shows a
low velocity expansion and is responsible for the central narrowest peak, and a double
elliptical cavity, discovered in the $J$=2--1 high sensitivity maps made by \citet{sc04}. As we 
see in the subsequent sections, the four nebular components are needed and must be included in the
nebula model to simultaneously reproduce  the observed spectral profiles. 

Although the profiles differ significantly, there are common trends that can be seen in Fig. 1. For
instance, the halo is the dominant component emitting in the low-$J$ transitions ($J$=2--1 and
$J$=1--0), whereas the high-velocity wings become progressively dominant as the level energy
increases, which are spectacular in the $J$=16--15 line. Another peculiarity of these $^{12}$CO lines is
the absorption features found in the $J$=2--1 and $J$=1--0 at velocities [--60,--40 \kms].
This has been  seen in other molecular lines like HCN ($J$=1--0) or HCO$^{+}$($J$=1--0) \citep{neri92,pardo04,sc04-b}. These
blueshifted features appear as a result of the continuum absorption and line self-absorption by cooler gas 
that is located in the outer envelope of CRL\,618. The absorption components are also observed 
in $^{13}$CO, as mainly seen in the $J$=1--0.
Again, there is a blueshifted feature at velocities of [--64,--52 \kms] that is believed to
arise from the outer envelope of CRL\,618, and another redshifted absorption component visible at  [0--10
\kms], which, as seen in the HCO$^{+}$ data, may not come from CRL\,618, since it coincides with
the velocity of the foreground gas of the Taurus dark cloud complex (3.5 \kms)
\citep{cerni89}. 

We also note that the central component of the profiles becomes significantly narrower in
high-$J$ transitions. This is due to the low velocity of the central dense core of the
nebula, where we can expect a high excitation.

Another common characteristic of the CO profiles is the clear asymmetries found in the spectral profiles.
 For instance, the  HIFI $^{12}$CO spectra show a high-velocity blue wing that is more intense than the red
one, which may be due to real asymmetries in the nebula. Also, the fast increase in
the wing/core ratio with the energy of the levels shows that the temperature of the fast outflow
may be higher than the values reported in previuos model fittings of the nebula (see next Section).
 In the $^{12}$CO data, especially in the HIFI spectra, we also note that the substraction of the
baselines is less accurate because of either the difficulty in selecting the 'complete' emission of
the corresponding line or the contamination due to emission from other molecules
\citep{bujarra12}.
In other cases as in the $J$=16--15 $^{13}$CO line, the baseline substraction is not accurate
enough due to the lack of spectral coverage in the observed frequency settings and the possible
presence of standing waves, which are often conspicuous at these high frequencies.

The $^{13}$CO data  show a core/ratio increase with $J$ but present weaker
wings. The $^{13}$CO/$^{12}$CO ratio is also much lower than 1, meaning that the $^{13}$CO lines
are optically thin. This idea is supported by the morphology of the line profiles. The profiles of
the low-$J$ $^{13}$CO lines (Fig. 1) are not typically parabolic as those found in optically thick
emission (see $^{12}$CO  in Fig. 1) but show the  complex kinematics of the inner parts of the
envelope, which result in profiles with a flat-topped line with spikes. As in the $^{12}$CO
lines, there are also asymmetries in $^{13}$CO; for instance,  the red-shifted region of 
the bipolar outflow is stronger and reaches up to $V$\,$\sim$\,40 \kms\ in the $J$=6--5 
and more clearly in the $J$=2--1 profile. 

A general description of the nebula model, a more detailed analysis and comparison between the
CO observed line profiles, and the results of the model fitting are given in the subsequent
sections.

\section{The nebula model}

We have used a numerical code that calculates  the excitation and emissivity of the
\doce\ and \trece\ rotational transitions throughout the various nebula components. The code
uses the {\em LVG} approximation to calculate the level populations of CO in a large number of points
in the nebula. The corresponding positions are defined by the distance of each point to the nebula
symmetry axis and to its equator. Then, the derived populations are used to calculate the emission 
and absorption coefficients to solve the standard radiative transfer equation and to determine 
the brightness distribution along a high number of lines of sight. Finally, the resulting brightness
 is convolved with the telescope beam (IRAM 30\,m or HIFI; see Table\ref{tab1}), which is described 
 by a Gaussian function, to obtain the line profiles in units of main beam temperature ($T_{\rm mb}$). 
This allow us to compare the model predictions with our HIFI observations. 

The {\em LVG} approximation can be used to calculate the excitation of a molecule when there is a sufficient
velocity gradient in the molecular cloud under study. Points separated a certain distance do not 
interact radiatively, and therefore only the local physical properties become relevant, which results 
in a considerable simplification of the calculations. In our case, this theoretical approach is
fully satisfied, since CRL\,618 is known to present high-velocity gradients. We note that this formalism yields
a good approximation to the excitation state, even in cases when the large velocity gradient
requirement is scarcely satisfied \citep{bujarra13}. In this paper, the authors present a wide 
discussion of the validity of this {\em LVG} approximation for a variety of conditions and conclude that
the calculations of the molecular excitation of CO are still accurate even in cases 
where the gradient of the velocity is low, which is not very dependent on the theoretical treatment of 
the radiative transfer. With this in mind, we have used the proper velocity gradient 
($\epsilon$\,=\,$\frac{{\rm d\,ln} V}{{\rm d\,ln} r}$) for each molecular component. For the dense core, 
ellipses, and outflow, $\epsilon$\,=\,1, and for the outer halo, $\epsilon$\,=\,0.1. In either case, we have 
checked that the calculations are consistent even if  we assume $\epsilon$\,=\,1 for the halo.

We have developed a new code that is based on \citet{bujarra97} but with significant improvements. Our
theoretical model performs more complex calculations of the excitation of both molecules, \doce\ and \trece.
 Since we are fitting the low-$J$ transitions and the high-$J$ lines, which are characterized by 
 high radiative probabilities, we cannot assume  that the energy levels are populated under $LTE$ in that case, 
 and therefore require  the $LVG$ approximation. To solve the statistical equilibrium equations, we have taken 
 the collisional coefficients  from the Leiden Atomic and Molecular Database \citep{schoier,yang}. In our 
 calculations, we have used collisions with ortho- and para-H$_{2}$, for which we have adopted an abundance ratio of 3. 
  As an additional test, we have performed  calculations considering the rotational levels within the $\nu$\,=\,1 
 vibrationally excited state and the effects of their presence in the populations of the $\nu$\,=\,0 levels. 
 In these calculations, we have considered $\Delta\nu$\,=\,1 collisional and radiative transitions 
 and the IR emission at 4.7 $\mu$m (the wavelength of the $\Delta\nu$\,=\,1 transitions) of CRL 618, assuming 
 that it comes from a small region in the center of the nebula. In all cases, these effects were found to 
 be negligible, so we conclude that further refinement of the treatment of the pumping via vibrational 
 states is not necessary.

The molecular structure of the nebula is based on the one proposed by SC04. 
Some general assumptions in our nebula model include (see Table~\ref{tab2} and discussions in SC04) (1)
a distance to the source of 900 pc, (2) an inclination of the source with respect to the plane of
the sky of 32\arcdeg, (3) a local dispersion velocity of 2 \kms, (4) relative abundances of
$X$(\doce)\,=\,3 10$^{-4}$ and $X$(\trece)\,=\,1.2 10$^{-5}$, and (5) symmetry with respect to the nebula axis
and equatorial plane. For the central HII region, we have used the same parameters as SC04: an
angular size of 8$\times$4 10$^{15}$ cm and a brightness temperature of 350 K, 
 which have been derived from their 1\,mm radio-continuum interferometric maps. We assumed that the free-free brightness varies with the frequency as $\sim$\,1/$\nu^{2}$. This is usually adopted for optically thin continuum emission in mm- and 
submm-waves. In all calculations, we find that the predicted line profiles depend slightly on the assumed HII region properties. 
At higher frequencies ($\sim$ 690 GHz, 1150 GHz, and 1800 GHz), we have also considered the contribution
of the dust continuum, which is known to be dominant \citep{wyrowski,lee}. To avoid extra parameters, we assumed that the dust brightness distribution is compact and similar to that of the free-free continuum, except for the observed values of the total flux.
 In any case, we find that the dust emission has a negligible effect in the excitation and profiles of these 
 high-frequency CO lines.  We have also assumed that the molecular abundance within the HII region is very low.

The geometrical parameters, velocity field, temperature law, and density distribution of the
different molecular components used in the CRL\,618 model are summarized in Table~\ref{tab2}. 
As mentioned, we have followed SC04 in the general description of the nebula by keeping almost exactly the same
nebular structure and kinematics, since that nebula model describes  the high-resolution
mm-wave maps of $^{12}$CO $J$=2--1 and HC$_{3}$N $J$=25--24, in which the main gas components can be
directly identified. However, the excitation conditions of the warm gas cannot be accurately derived from those
transitions, as already discussed. We have accordingly modified the density and temperature in the
various nebular components to fit our high-$J$ transitions and to  reproduce the observational
data at low-$J$. Several nebula components are identified: a
large and diffuse spherical halo radially expanding with an outer velocity of $\sim$\,16 \kms, a
compact and dense central core also expanding radially with a constant temperature of 120 K and
a density decreasing rapidly with the distance to the star $r$, a double empty shell radially
expanding at low velocity with a constant temperature of 80 K, and finally, a high-velocity 
outflow with a cylindrical shape except for the inner regions (see Fig. 2 and Table~\ref{tab2})
where it is conical. The outer region of the high-velocity outflow is better reproduced by assuming
pure axial expansion, whereas the inner parts are radially expanding. The temperature law is the same for the
entire outflow, but the density decreases with distance slightly faster in the cylindrical region
(see details in Table~\ref{tab2}). We recall that the code determines the excitation for both
molecules, \doce\ and \trece, and that the only difference in the calculations is their
relative abundance to H$_2$ and their molecular constants.

In general, the predictions from our code using this nebula structure satisfactorily 
reproduce the observed CO profiles as we discuss in the next section,.  


\begin{table*}
\caption{Molecular components of the model for CRL\,618.}
\begin{tabular}{lr@{\,=\,}ll@{}cccc}
\hline 
\hline\\[-8pt]
Nebula    & \multicolumn{2}{c}{Size} & &Radial Velocity & Temperature & Density & Mass\\
component & \multicolumn{2}{c}{(cm)}     & &$V_{\rm{exp}}$ (\kms) & $T$\,(K) & $\rho \,(\rm{cm^{-3}}$) & $M$\,($M_{\odot}$)\\[2pt]
\hline
\hline\\[-5pt]
Halo ............... & $R_{\rm{in}}$ & 1.6 $\times$ 10$^{16}$ & & 7 + 10 [1 --
($R_{\rm{in}}$/$r$)]$^{1/2}$& 115 (10$^{16}/r$) + 5
&1.2\,$\times \,10^{6}$ (10$^{16}$/$r$)$^{2}$& 0.24\\

&$R_{\rm{out}}$& 10$^{17}$&&&\\[4pt]

Ellipses .......... &$M$=major axis&8.4 $\times$ 10$^{16}$&&  10 [1 + ($h$ / $M$)]&80& [1.1
-- 1.05 ($h$/$M$)]10$^{6}$& 0.04\\
&minor axis &4 $\times$ 10$^{16}$&&\\
&thickness &2.6 $\times$ 10$^{15}$&&&\\
&center &$\pm$4.2 $\times$ 10$^{16}$&&\\[4pt]

Fast outflow ... &$R_{\rm{out}}$&0.9 $\times$ 10$^{16}$&(a)& \,45 + 255
($h$--10$^{16}$)/($h_{max}$ -- 10$^{16}$)& 300 (10$^{16}$/$h$) +
5&5.5 $\times$ 10$^{6}$ (10$^{16}$/$r$)$^{2}$ & 0.07\\[2pt]
&$h_{\rm{max}}$&2.8 $\times$ 10$^{16}$& (b)&45 [1 + ($h$--10$^{16}) / (10^{16})$] &300 (10$^{16}$/$h$)
+ 5 &5.5 $\times$ 10$^{6}$ (10$^{16}$/$r$)\\[4pt]

Dense core ..... &$R_{\rm{out}}$&1.6 $\times$ 10$^{16}$& &7 ($r$ / $R_{\rm{out}}$)&120
&2.7 $\times$ 10$^{6}$ (10$^{16}$/$r$)$^{3}$& 0.05\\[2pt]

\hline
\hline
\end{tabular}
\label{tab2}
\\[2pt]

$r$ is the radial distance to the central star. \\
$h$ is the distance to  the equator. \\
(a) and (b) are the values for the  cylindrical and  conical regions of the high-velocity outflow.

\end{table*}


\begin{figure}[]
\centering
\includegraphics[width=0.5\textwidth]{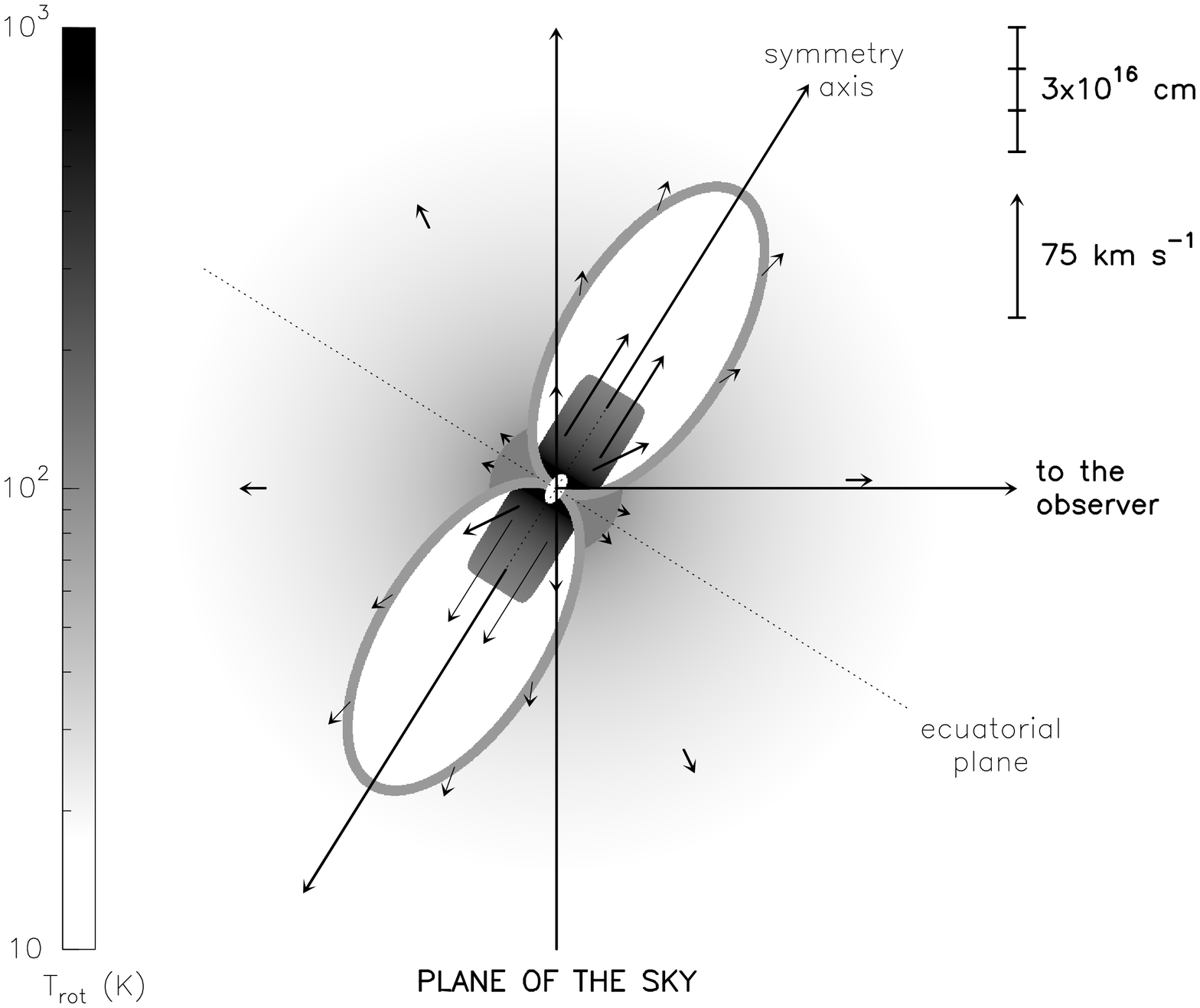}

\vspace{.5cm}
\includegraphics[width=0.5\textwidth]{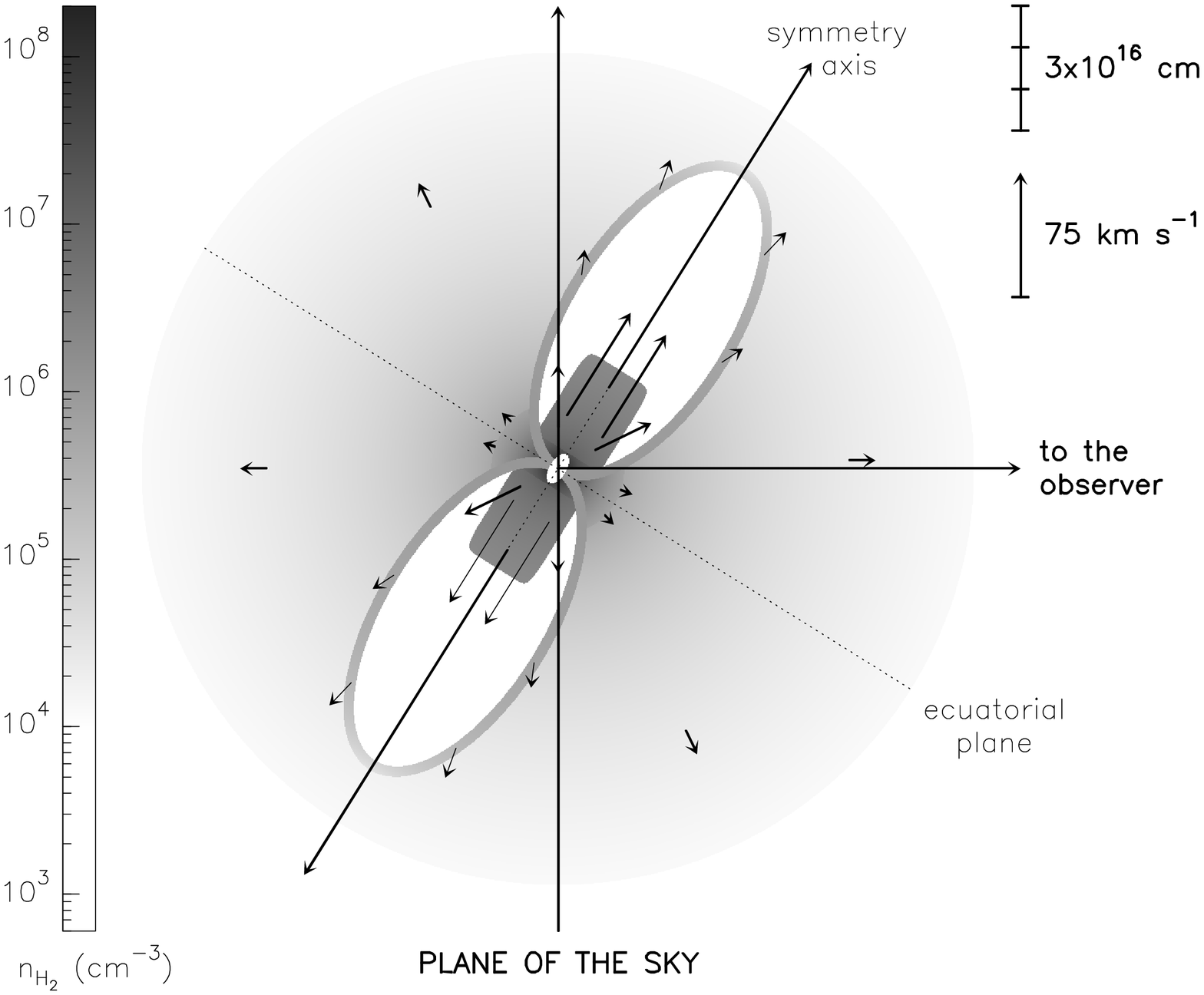}
\caption{Representation of the temperature law and density distribution of the derived
molecular components of CRL\,618 from our theoretical model. 
The plot also shows the velocity field (arrows) and the compact HII region (elliptical central shell).
 }
\label{fig2}
\end{figure}

\section{Discussion}

As mentioned, we have followed the model made by SC04 by using a similar nebular structure.
We have introduced moderate variations in the geometry and velocity field and more substantial
changes in the temperature and density laws, which characterize the different nebular components that
allow a better fit of the observed high-$J$ CO profiles. In particular, the main differences,
when compared to SC04, are found in two components: the high-velocity outflow and the dense central core. 
Of course, we must keep in mind that we are analyzing molecular lines that require different
excitation conditions (for example, the lowest and highest rotational transitions which are
separated by more than 700 K) and therefore, are tracing distinct regions of the nebula, which lead
to a more accurate determination of physical parameters like the density and temperature than 
in SC04. 

We have compared our observational data with the predictions of the model summarized in Sect. 4.
We have derived the best nebula model by fitting the observed CO profiles (see  Table~\ref{tab2}).
Our results are shown in Fig. 1. We can see that our predictions for the best nebula model 
 adequately  reproduce the observed spectral profiles for both molecules, \doce\ and \trece, in CRL\,618. 
To get to this satisfactory fitting, we have used  the same
nebular structure and identical physical conditions of those used by SC04 as a first step. With this approach, we
were able to reproduce the $J$=1--0 and $J$=2--1 lines but not the high excitation ones. To do the
latter, we needed to change the physical parameters of the different components. 
The general properties of the halo and the double cavity are not changed significantly 
(see Table~\ref{tab1} in SC04 and our Table~\ref{tab2}). For instance, we have introduced a small velocity 
gradient of the gas in the extended halo from 7 up to 16 \kms, which is almost the same value 
derived in SC04.  
The temperature law is similar with the halo slightly being hotter. The density distribution is  similar too,
although the inner regions of the halo are denser than in SC04. 
On the other hand, the double empty cavity that extends
along the nebula axis (named ellipses in Table~\ref{tab2}) also shows a maximum velocity of 20 
\kms, and temperature and density laws that are compatible to those used in  SC04. The ellipses
in this case are cooler and less dense. This result confirms the predictions made in
\citet{bujarra10}.
As we can see, the small differences in these two nebular components must be included in
the model to simultaneously fit the low-$J$ emissions (which mainly comes from the halo) and the
emission from the ellipses which, although its contribution to the profile is not very important as
shown by the observations, is seen in both the $J$=10--9 and $J$=6--5 spectra of $^{12}$CO and
$^{13}$CO and in the $J$=16--15 $^{12}$CO line. 

In contrast, the physical conditions and properties of the other two components, the bipolar outflow 
and the compact central component, have been revised more substantially in our model. We discuss 
these changes below.

1. \emph{The high-velocity bipolar outflow}:  A first approach to model the high CO emission was performed by
\citet{bujarra10}. These authors made a preliminary fitting of the $J$=16--15 $^{12}$CO line observed
in CRL\,618. The main conclusion they found was that the excitation conditions to reproduce the
intensity measured in the outflow were underestimated -- a result that we have confirmed in our model
predictions. In particular, the velocity and density distributions do not differ  much from the
original model but the temperature laws have  changed. In the entire outflow we have used the same temperature law
(see Table~\ref{tab2}); the temperature decreases with the distance to the star, being $\sim$\,300 K
at a typical distance of $r$=10$^{16}$ cm. This relatively high temperature is needed to
reproduce the evident contribution of the outflow to the central observed features  (see $J$=16--15, 10--9,
6--5 in Fig.\,1) and the high-velocity wings seen in the different high frequency spectral profiles. 
Our estimation of the temperature in the outflow of CRL\,618 is higher than 
the values typically derived in other PPNe, which are usually lower: $\sim$\,100 K in CRL\,2688 
\citep{cox00,bujarra12} and $\sim$\,60 K in Frosty Leo \citep{castro05} and IRAS\,17436+5003 \citep{bujarra12}. 
However, these values are consistent with the derived averaged density and time, since the acceleration
of the outflow gas took place \citep{bujarra12}. Using the velocity gradient and size of the outflow in 
Table~\ref{tab2}, we derive a kinematic age of the outflow in CRL\,618 of $\sim$\,70 yr in the inner
regions or close to the central star and $\sim$\,30 yr in the outermost regions. The gas in the outflow of CRL\,618 is then 
an example of the very recent acceleration and heating of the passage of shocks that is still cooling down.

We note that the observed asymmetry between the red and the blue line wings is not
reproduced in our calculations. In general, the blue wing is  more intense than the
red one in the high-$J$ transitions, while the red wing is stronger in the low-$J$ lines,
particularly for \trece. As mentioned in Sect.\,3, opacity effects in opaque lines can yield 
weaker blue wings, due to absorption by outer cooler layers approaching the observer.  This phenomenon, 
which is widely observed in molecular lines, cannot explain the empirical features in our case, 
which must be due to actual asymmetries of the nebular properties with respect to the equator. 
We think that our observations reveal the presence of slightly higher temperatures in the 
approaching gas, which could explain the intense high-$J$ blue wings, and a higher mass in the receding gas, which would be responsible for 
the stronger red wing found in low-$J$ \trece\ lines. However, our model cannot account for these features, 
since we assume symmetry with respect to the equator for simplicity, and therefore the calculations 
always give  an average of the emission of both wings.

2. \emph{The central core}: This component was proposed by SC04 to explain the compact and intense
emission found in their two central channels at $V$=[--18,--24 \kms].  
In the $J$=16--15, 10--9 transitions of $^{12}$CO and $^{13}$CO, it is clear from Fig. 1 that
there is a central feature that is actually emitting at the velocities seen by SC04. To
properly fit the low-CO emission and the high-CO, we have to introduce again small
differences to the physical laws used in SC04. For instance, the fitting is improved if the velocity
of this component is decreased to typical values of $\le$ 7 \kms\ while keeping its strong
gradient to reproduce the triangular shape observed in the profiles. In addition, a constant
temperature of 120 K and a less dense gas in the core are needed to fit the profiles satisfactorily
(see Figs. 1 and 2, and Table~\ref{tab2}).

Other assumptions in our model concern the relative abundances of $^{12}$CO and $^{13}$CO. We have 
used an abundance ratio, \doce/\trece, equal to 25. 
This ratio agrees with the values derived in C-rich circumstellar AGB
envelopes and PPNe \citep[see][]{kahane,bujarra94,bujarra01,bujarra10}. We have also assumed that
both values are constant throughout the nebula.

Using our best possible model for CRL\,618, we derive a total molecular mass of $M$\,$\sim$\,0.4 $M_{\odot}$.
Most of the calculated mass is contained in the halo and in the bipolar outflow with values of 0.24 $M_{\odot}$ and 0.07 $M_{\odot}$ respectively. The mass of the dense core is 0.05 $M_{\odot}$;
 the ellipses are less massive with a mass of 0.04  $M_{\odot}$. 
 The kinetic energy and total momentum of the nebula
are 9.1 $\times$ 10$^{45}$ erg and 1.9 $\times$ 10$^{39}$ g cm s$^{-1}$. The outflow carries most of the
momentum with a value of $\sim$\,1.1  $\times$ 10$^{39}$ g cm s$^{-1}$.
These values are compatible with the known previous estimations
of the masses and linear momentums by \citet{bujarra01} and SC04, when we consider the differences
 in the methods and/or parameters used. In the first case, the authors used a simple nebula model, a distance of 1700 pc 
 (instead of 900 pc), and a very low rotational temperature (15 K for the outflow and 25 K in the slow component) 
 when compared to our derived mean values ($\sim$\,300 K and $\sim$\,100 K, respectively). 
In SC04, the masses of the core and outflow, and hence the momentums, are larger than our values. 
This difference is explained due to the coefficients of the density laws assumed in the two components. Our
coefficients in the density laws are lower than those derived in SC04. There is also another difference
with SC04: We have used a higher abundance of \doce. This means that we expect lower densities throughout the nebula 
for similar line intensities, since there are more molecules emitting at the different frequencies. 
Finally, if we compare the momentum of the outflow with what can be supplied
by the radiation pressure ($\sim$\,3.5\,$\times$\,10$^{34}$ g cm\,s$^{-1}$\,yr$^{-1}$), which is a result that does not
depend on the assumed distance, we obtain $P/(L/c)$ $\sim$\,0.3\,$\times$\,10$^{5}$\,yr. This high value is comparable to the one 
given by SC04 ($\sim$\,10$^{5}$ yr), which, as explained by the authors, confirms that radiation pressure
cannot drive the observed bipolar outflows seen in CRL\,618.

In general, the observed spectral profiles are well reproduced by our model. We find that the errors
in the intensities derived from the fitg range between 15 and 20$\%$, except for the $^{13}$CO
$J$=16--15 line where the error in the determination of the total flux is 50$\%$. These scale
factors are then comparable to the errors measured in the amplitude calibration of the processed data
(see Table~\ref{tab1}), except for the $^{13}$CO $J$=16--15 where the detection is
within the observing limits.

The nebular structure and velocity field are independent of the physical conditions
used in the different components, since they have been determined directly from the available data.
The determination of the temperature and density is usually more uncertain
because it depends on the individual properties of each molecular component, where the opacity of
the CO emission changes across the nebula. As mentioned previously, a proper study of 
those parameters also requires the observations of high excitation lines,
and therefore,  we have been able to derive reliable values of the temperature and density 
throughout the molecular envelope thanks to our Herschel observations.

\section{Conclusions}

We present observations of the FIR/sub-mm molecular lines of $^{12}$CO and $^{13}$CO using 
Herschel/HIFI in the protoplanetary nebula CRL\,618.  The $J$=16--10, 10--9, and 6--5 lines of both 
isotopomers have been detected. We have improved the reduction and, particularly, the calibration of 
these observational data, which were previously reported. 

 We have used a nebula model that calculates the excitation of the CO transitions through the 
 molecular components of the nebula.  The model includes accurate calculations of the molecular 
 excitation and radiative transfer in a complex structure. Our theoretical approach also allows a good 
 determination of the temperature thanks to our modeling of the high excitation transitions. 
 For the  first time, a spatio-kinematical model has been used to fit all the observed low- and
 high- excitation CO lines simultaneously, and therefore, we have been able to better 
 constrain and study  relevant parameters like the density and temperature across the nebula. The main 
 conclusions derived  from this study are the following:

\begin{enumerate}
   
\item We confirm the several molecular components previously identified in
CRL\,618: a spherical halo, a dense core, a double empty shell, and a high-velocity outflow. 
The overall properties of the halo and the double shell do not differ 
significantly from previous studies. The halo, which is probably the remnant of the AGB circumstellar envelope,
contributes to the line profiles mainly in the low-$J$ CO transitions.
  The double empty shell, on the other hand, extends along the nebula axis and its 
  contribution is less important at high-$J$ emission. 
 In contrast, the compact central core and the high-velocity bipolar outflow are the dominant components at higher
 excitations. 
 
\item To obtain the best possible fit to the low-$J$ and particularly the high-$J$ CO profiles, we
have found that important changes in the physical conditions in the bipolar outflow and core must be considered: 
 
\begin{itemize}

\item The fast bipolar outflow, which axially expands from $\sim$\,45 to 300 \kms\ with a typical density of
5.5 $\times$ 10$^{6}$ cm$^{-3}$, is characterized by a relatively higher temperature than previously 
estimated; the temperature decreases with the distance to the central star, being $\sim$\,300 K at a
typical distance of 10$^{16}$ cm. This temperature is needed to explain the high-velocity wings 
seen in the spectral profiles, which become dominant when the energy level increases and which are
spectacular in the $J$=16--15 $^{12}$CO line. This temperature for the fast gas is also higher
than what is found in PPNe. Therefore, the outflow in CRL\,618 may be the result of very
recent acceleration of shocked gas that has not cooled down.
      
\item The molecular emission arising from the compact central core is clearly observed in the low- and 
high-$J$ transitions. The core emission is the narrow central feature seen in the high-$J$ line profiles.
 Therefore, its temperature is higher,  and its expansion velocity is lower than estimated. 
In particular, the temperature has been increased to 120 K, and the radial expansion velocity 
 has been decreased to typical values of $\sim$\,5 \kms, while keeping its strong gradient. 
 This molecular component may then represent the ejected material from the central star during the ultimate 
 AGB phase, at around 500 yr, where AGB stars experiment a higher mass-loss rate and, at the same time, a 
 drop in the expansion velocity.
 \end{itemize}
 
   \end{enumerate}

\begin{acknowledgements}

The authors acknowledge the useful comments of the referee, which helped to improve the paper.
      
\end{acknowledgements}

\bibliographystyle{aa}
\bibliography{biblio}

\end{document}